\documentclass{mem}
\usepackage{natbib}\usepackage{txfonts}\usepackage{balance}
\usepackage{graphicx}
\usepackage[a4paper]{hyperref}

\begin{document}

\title{Supercomputing and stellar dynamics}

   \subtitle{}

\author{
R. \,Capuzzo-Dolcetta
          }

  \offprints{R. Capuzzo--Dolcetta}

\institute{
Dipartimento di Fisica --
Sapienza, Universit\`a di Roma, P.le A. Moro 2, 
I-00185 Roma, Italy
\email{roberto.capuzzodolcetta@uniroma1.it}
}

\authorrunning{Capuzzo--Dolcetta}

\titlerunning{Supercomputing and stellar dynamics}

\abstract{
In this paper I will outline some of the aspects and problems of
modern celestial mechanics and stellar dynamics, in the context of the 
quickly growing computing facilities. I will point the attention
on the great advantages in using, for astrophysical simulations, the modern,
fast and cheap Graphic Processing Units (GPUs) acting as true supercomputers.
Finally, I present and discuss some characteristics and performances 
of a new {\it double-parallel} code exploiting the joint power of multicore 
CPUs and GPUs.

\keywords{Celestial mechanics-- Stellar dynamics-- 
Methods: N-body simulations-- Supercomputing}
}
\maketitle{}

\section{Introduction}

The role of gravity in physics is, of course, fundamental. Anyway, this role
is completely different in earth physics respect to that in astrophysics.\\
In terrestrial physics gravity is not too difficult to be accounted for,
because it simply acts as an external constant field to add to other 
more complicated interaction among the constituents of the
system under study. In other words, physical systems on earth are not 
self-gravitating, and this implies an enormous simplification. In an
astrophysical context, things are different: astronomical objects are 
self-gravitating. Their shape, volume and dynamics are
determined mainly by self-gravity, which acts, often, in conjunction with 
the external gravity due to the presence of either outer bodies or
general potential where the object is embedded in. External gravity
determines the orbit of the astronomical body, and influences its shape, 
at least in its outskirts, by mean of tidal interactions, as well.

A simple quantitative parameter to measure the role of self-gravity to the whole
energetics of a given system is the ratio, $\alpha$, between the
self-gravitation
energy of the system and the energy given by the external gravitation field
where the system is embedded in.
For a typical terrestrial system like, e.g., the Garda lake $\alpha \simeq 10^{-8}$, 
while for two, quite different, astronomical systems (a typical globular cluster
in a 
galaxy and a typical galaxy in a galaxy cluster) $\alpha \simeq 10^{-2}$: 
a million times greater. Apart from the other, obvious, differences (a lake is
composed by a liquid, where the collisional time scale is negligible respect to any other
time scale in the system, while the globular cluster and the galaxy are composed by stars moving
in volumes such that the collisional 2-body time scale is comparable,
in the case of globular cluster, or much longer, in the case of galaxy, to the
system orbital time and age), it is clear that while the lake molecules mutual
gravitational interactions are negligible respect to the external field, this is
not the case for the stars in globular clusters or galaxies.

\section{$N$-body systems in astrophysics}

As stated above, self-gravity cannot be neglected when studying the
physics of astronomical objects. This makes theoretical astrophysics a 
hard field: the dynamics of astrophysical systems is intrinsecally difficult 
to be studied, even in newtonian approximation, because of the 
{\it double divergence} of the, simple, two-body interaction potential, 
$U_{ij}\propto 1/r_{ij}$, where $r_{ij}$ is the euclidean distance between 
the $i$ and $j$ particle,\\
$r_{ij}=\sqrt{(x_i-x_j)^2+(y_i-y_j)^2+(z_i-z_j)^2}$. {\it Ultra-violet}
divergence corresponds to very close encounters, {\it infra-red} divergence to that the gravitational interaction
never vanishes. 
These divergences introduce a multiplicity of time scales (Aarseth 1985) and
make impossible to rely on statistical mechanics and/or to non-perturbative 
methods, as often done in other particle-systems physics.
Actually, the newtonian $N$-body dynamics is mathematically represented by the
system of $N$ second-order differential equations

\begin{equation}
\label{system}
\left \{ \begin{array}{ll}
\ddot{\textbf{r}_i}=G\sum^N_{j=1, j \not= i}
\frac{m_{j}}{r_{ij}^{3}}\,
(\textbf{r}_{j}-\textbf{r}_{i}),\\
\dot{\textbf{r}_i}(0)=\dot{\textbf{r}_i}_{0},\\
\textbf{r}_i(0)=\textbf{r}_{i0},\\
(i=1,2, \dots ,N).
\end{array}
\right.
\end{equation}

\noindent This dynamical system is characterized by: (i) $O(N^2)$ complexity,
(ii) being far from linearity, (iii) having few constraints in the phase-space.
Sundman (1912) showed (without winning the King Oscar II 
Prize, already awarded to H. Poincar\'e...) that for the three-body 
problem there is a series solution for the coordinates in powers of 
$t^{1/3}$ convergent for all $t$, except initial data
which correspond to zero angular momentum. 
This result was generalized to any $N$ just in relatively recent times 
by Wang (1991). 
Anyway, the power series solutions are so slow in convergence
to be useless for practical use. This means that the gravitational $N$-body
problem must be attacked numerically.\\
The difficulties in doing this are, both, {\it theoretical} and {\it
practical}. On the {\it theoretical} point of view, one has to face with the
chaotic behaviour of the nonlinear system which is related to the extreme 
sensitivity of the system's differential equations to the initial conditions: 
a very small initial difference may result in an enormous change in the 
long-term behaviour of the system. Celestial dynamics gives, indeed, one of the
oldest examples of chaos in physics. This problem is almost
unsolvable; it may be just kept under some control by using sophisticated, high
order time integration algorithms. On the {\it practical} side, the (obvious)
greatest complication to face is the due to the infrared (large scale) divergence, that
implies the need of computing all the $\propto N^2$ force interactions between
the pairs in the systems. This results in an extremely demanding computational
task, when $N$ is large (see Table I).
We will now discuss some of the problems arising when dealing with the
numerical study of the evolution of self-gravitating systems over the
astronomical range of $N$.

\begin{table*}[]
\begin{center}
\caption{Some typical astronomical systems, with their star number (N), 
number of floating point operations needed for the force evaluations 
in a single system configuration (n$_f$) and CPU time  
required to the n$_f$ operations  by a single processor of 1 Gflops speed
(t$_{CPU}$, in seconds).
Note that $1.8\times 10^{14}$ sec $\simeq 5.7$ Myr!}
\label{cputimes}
\begin{tabular}{|l|c|c|c|} \hline \hline
system  & $N$    & $n_f$ & t$_{CPU}$\\ \hline
Open cluster & $1000$  & $1.5\times 10^7$ & $0.02$  \\ \hline
Globular cluster & $10^5$  & $1.5\times 10^{11}$ & $180$\\ \hline
Galaxy & $10^{11}$ & $1.5\times  10^{23}$ & $1.8\times  10^{14}$ \\ \hline
\hline
\end{tabular}
\end{center}
\end{table*}

\section{Small- and Large- $N$ systems: from celestial mechanics to stellar 
dynamics}

On the small- $N$ side ($N \leq 10$, example: solar system) the problem is not
that of enormous CPU time consumption, for the number of pairs is small, but,
rather, that of the need of an enormous precision. This to keep the round-off
error within acceptable bounds when integrating over many orbital times. 
In the case of few bodies, reliable investigations cannot accept
the point mass scheme (for instance, the Sun potential requires a multipole
expansion) and high precision codes are compulsory. Pair force evaluation is
computationally cheap due to the low number of pairs; 
on the other side, even very small round-off errors increase secularly, time
step by time step, making high order symplectic integration algorithms 
unavoidable. The need is: a fast computer, able to handle with motion 
integration over a very extended time and able to evaluate forces with 
enormous precision.\\ 
We do not speak any further of the few body regime, which is the realm of modern
celestial mechanics and space dynamics, but go to say something on the problem
of intermediate- and large-N-body systems, task which is typical of the modern stellar dynamics, instead.
Force computation by pairs is computationally expensive, the mostly demanding
part being the evaluation of the distance $r_{ij}$ between the generic $i$ and $j$
particle. It requires the computation of a square root which, still with modern
computers, is based on ancient methods among which the Erone's method, the Bombelli's method 
and the Newton-Raphson numerical solution of the quadratic equation $x^2- r_{ij}^2=0$.
In any case, the single pair force evaluation requires about 30 floating 
point operations; this means that in an N-body system, 
$n_f = 30\times N(N-1)/2$ floating point operations are required. 
A single processor (PE) with a speed of 1 Gflops would compute the single pair
force in $\sim 3\times 10^{-8}$ sec. Consequently, the whole N star forces
would require the time indicated in Table I for their evaluation at every time
step. Clearly, the task of following numerically the long term evolution of a
large- N-body system by a program based on direct summation of pair forces is very far
out of the capability even of the most performing computers.
Actually, the profiling of any computer code to integrate N-body evolution
indicates that about $70\%$ of the CPU time is spent in force evaluation. 

What strategies must be used, then?

The most natural way to attack the problem is a proper combination of the
following ingredients: 
(i) simplification of the interaction force calculation; (ii) reduction of the
number of times that the forces have to be evaluated, by a proper
variation of the time step both in space and in time; (iii) use of the most
powerful (parallel) computers available.\\
Points (i) and (ii) require a deep effort of numerical analysis, point (iii) 
requires the solution of the, not easy, problem of parallelizing an N-body code. 

The simplification of force calculation may be done by means of the 
introduction of space grids, both for computing the large scale component 
of the gravitational force via the solution of the Poisson's equation (with 
Fast--Fourier codes, for example) and for the dynamic subdivision of the space
domain with a recursive, octal tree to take computational advantage by a multipole 
expansion of the interaction potential (approach first used by Barnes \& Hut 1986). 
These are two of the possibilities to reduce the particle-particle (PP) force
evaluation to a particle-mesh (PM) or particle-particle-particle-mesh (P3M) approach, 
with obvious computational advantages (see Hockney \& Eastwood 1988 for a general
discussion). 
In addition to the complications introduced in the computer code, a clear limit
of this procedure is the error introduced in the force evaluation,
which can be reduced, over the small scale, by keeping a direct PP force
evaluation for close neighbours. Point (ii), time stepping variation, relies
mainly on the use of individual (per particle) time steps.
Particles are advanced with a time step  proper to the
individual acceleration felt, allowing a reduction in highly
dynamical cases without stopping the overall calculation. Unfortunately,
individual time stepping requires careful implementation to guarantee 
synchronous integration and implies, often, a reduction of order of precision of the integration method. Finally, the
parallelization of gravitational codes (point (iii)) is difficult, because
gravity is such that the force on every particle depends on the position of all
the others. This makes non trivial a domain decomposition such to release a 
balanced computational weight to the various PEs of a
parallel machine. In this context, it is relevant noting that many active groups
of research chose to use ~\lq dedicated\rq~ parallel architectures, which act as
boosters of specific computations, like those of the distances between
particles. This is the road opened by the Japanese GRAPE group lead by Makino
(Makino 1991).
Another, intriguing, possibility is the use of Graphic Processing Units (GPUs)
as cheap alternatives to dedicated systems.
GPUs are used to speed up force computations and give high computing
performances at much lower costs, especially in cases where double precision 
is not required.
This is the choice explored in astrophysics first by S. Portegies Zwart and his dutch group (Portegies Zwart, Belleman \& Geldof 2007).  
Capuzzo--Dolcetta and collaborators in Italy (Capuzzo--Dolcetta, Maschietti \& Mastrobuono--Battisti 2009) have constructed a direct N-body code implementing 
sophisticated 2$^{nd}$ and 6$^{th}$ order symplectic time-integration and using as force evaluation accelerator 
a pair of brand new NVIDIA TESLA C1060 Graphic processing Units (GPUs) programmed by means of the native NVIDIA 
Compute Unified Device Architecture (CUDA, see \\
{\it www.nvidia.com/object/cuda\_home.html}).

\section{The NBSymple code} 

The code generates, first, the initial conditions for the $N$-body system, 
whose individual masses may be chosen by a given mass spectrum. The total 
mass of the system, $M$, is assumed as mass unit. For the sake of simplicity, aiming 
first at checking quality of integration and at performances testing, particles 
were given an initial spatially uniform distribution within a sphere of given 
(unitary) radius, $R$, with  velocities, also, uniformly distributed in direction and absolute values
and rescaled, in their magnitude, to reproduce a given value of the virial 
ratio. We remind that the virial ratio is defined as $Q=2K/|\Omega|$, 
where $K$ and $\Omega$ are, respectively, the system kinetic and 
potential energies; for a stationary system, $Q=1$. Note that the 
further assumption $G=1$ in the equations of motion implies that 
the ~\lq crossing\rq~ time $T=(GM)^{-1/2}R^{3/2}$ is the unit of time.

The code allow the introduction of a softening parameter ($\epsilon$) in the star-star 
interaction potential, usually taken as a fraction of the closest neighbour 
average distance. The pairwise forces are summed to the force due to the 
external field, which is accounted by an analytical expression for the 
Galactic potential as given by Allen \& Santillan (1991). In this latter work the 
authors consider the Galactic potential as given by three components: 
a bulge, a disk and a halo.
The bulge and the halo have a spherical symmetry, while the disk is 
axisymmetric.\\
Any kind of generalization to diferent sets of initial conditions and external potentials
is easy done by mean of appropriate external subroutines provided by the user.

\subsection{Time integration} 

It is well known that ordinary numerical methods for integrating Newtonian
equations of motions become dissipative and exhibit incorrect long term
behaviour. This is a
serious problem when facing $N$-body problems, particularly when studying their long term evolution.
One possibility is to use symplectic integrators.
Symplectic integrators are numerical integration schemes for Hamiltonian
systems, which conserve the symplectic two-form ${\rm d}{\bf p}\land 
{\rm d} {\bf q}$ exactly, so that
(${\bf q}(0),~{\bf p}(0))$ 
$\to({\bf q}(\tau),~{\bf p}(\tau))$ is a canonical transformation. The transformation is
characterized by time reversibility. 
If the integrator is not symplectic, the error of the total energy grows 
secularly, in general. 
Our code allows the choice of two different symplectic methods.
One is the simple, classic ~\lq leapfrog\rq~ method, which is second order 
accurate; the other is a more accurate sixth order explicit scheme 
whose coefficients are taken from the first column of the Table 1 of  
Kinoshita, Yoshida \& Nakai (1991), which leads to a time integration conserving 
energy much better than that with the other two possible sets of coefficients in the
Tableg. Of course, the 6-th order symplectic integrator is much slower than the leap 
frog, requiring 7 evaluations of force functions per time step, like, for instance, in a 6-th order Runge Kutta method).\\

\subsection{The computing platform}

The workstation used to test and run our NBSymple code has a 2 
Quad Core Intel Xeon processors, each running at $2.00GHz$, $4$GB DDR2 
RAM at $667MHz$ and two 
NVIDIA TESLA C1060 GPUs, connected to the host via two slots PCI-E 16x.
\\
NVIDIA TESLA C1060 has 240 processors, each of them has a clock of 1.296 GHz.

\section{Results}
Accurate testing of both the quality of the $N-$body system integration 
and of the computational efficiency of NBSymple is given in the Capuzzo--Dolcetta, 
Maschietti \& Mastrobuono--Battisti (2009) paper. In that paper, the various 
versions of the code are presented and discussed.
Some versions work in single-precision 
arithmetics (exploiting at best the GPU performances but not fully 
satisfactory in terms of the precision) and in both harwdware (slower, more precise) 
and software (faster, less precise) double-precision arithmetics.
The software double-precision is implemented following Goburov, 
Harfst \& Portegies Zwart (2009).\\
The NBSymple code has presently 5 versions, each labeled with an alphabetic letter from A
to E:

\begin{itemize}
\item NBSympleA: fully serial code running on a single Quad core processor;
\item NBSympleB: single-parallel code which uses Open Multi-Processing (OpenMP) directives, 
for both the $O(N^2)$ pairwise interactions and the $O(N)$ calculations (i.e. the time integration and evaluation of the Galactic component of the force on the system stars) 
over the double Quad core host;
\item NBSympleC: single-parallel code, where the ($O(N^2)$ all-pairs interactions 
calculations)are demanded to the NVIDIA TESLA C1060 GPU, using CUDA while all the 
remaining tasks are done by a single Quad core CPU; 
\item NBSympleD: double-parallel code, which again uses CUDA to evaluate the 
$O(N^2)$ portion of the code (as NBSympleC), while the $O(N)$ computations 
is parallelized sharing work between all
the eight cores of the host, using OpenMP, as NBSympleB;  
\item NBSympleE: single-parallel code that uses CUDA on one or two GPUs to
evaluate the total force over the system stars, i.e. both the all-pairs component 
and that due to the Galaxy.
\end{itemize}

\begin{figure*}[htbp]
\begin{center}
\includegraphics[width=\textwidth]{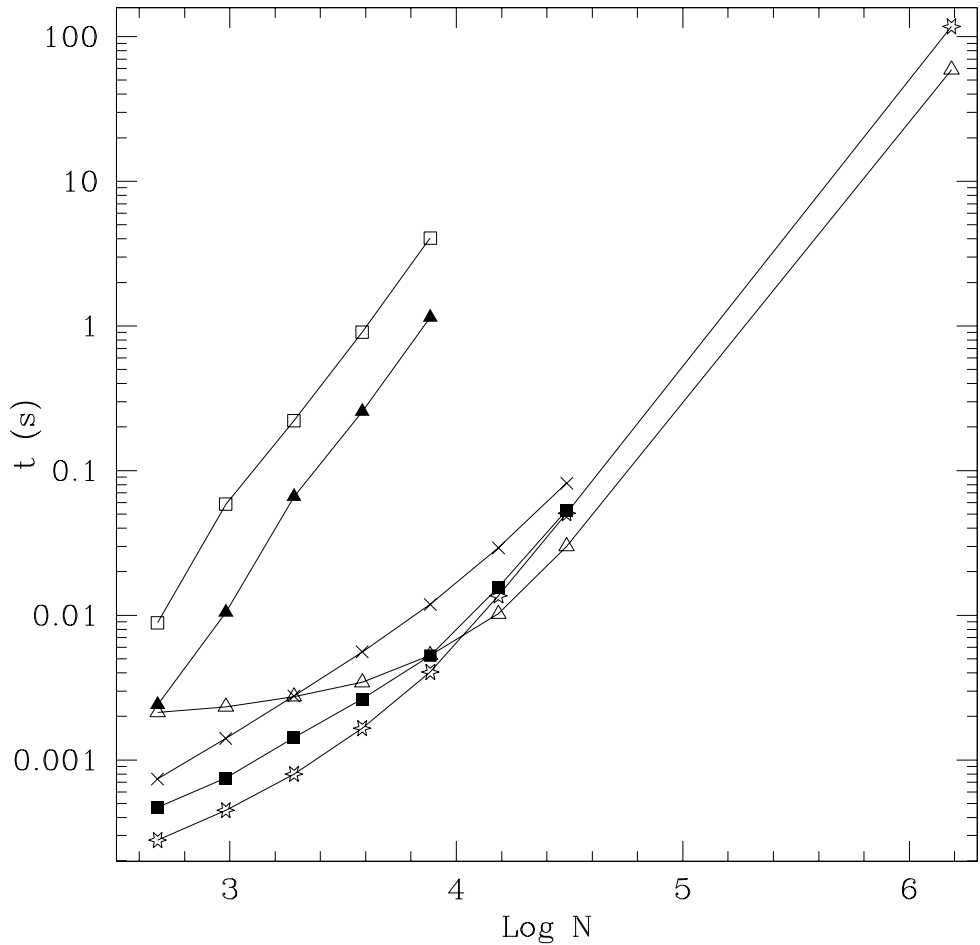}
\caption{The (averaged over $1000$ cycles) solar time (in seconds) 
spent by various versions of NBSymple to perform a single integration step, as a 
function of $N$. Line with empty squares: 
NBSympleA code. 
Line with filled triangles: NBSympleB. Line with crosses: NBSympleC. 
Line with filled squares: NBSympleD. Line with empty triangles: NBSympleE 
with a single GPU.
Line with stars: NBSympleE with two GPUs.}
\label{speedvsN} 
\end{center}
\end{figure*}

We emphasize that the pairwise interactions evaluation are developed, in the CUDA framework, following
mainly the Nyland, Harris \& Prins (2007) work.\\ 
Here I just present a figure (Fig. \ref{speedvsN}) showing a comparison of the 
time spent (in seconds) by various versions of the NBSymple code for a single time 
step integration of an $N-$body system as function of the number of bodies.

\begin{acknowledgements}
I am grateful to my collaborators D. Maschietti and A. Mastrobuono--Battisti,
whose help was fundamental in developing the code in the CUDA frame.
\end{acknowledgements}

\bibliographystyle{aa}

\end{document}